# Photoinduced surface plasmon control of ultrafast melting modes in Au nanorods

Eunyoung Park[1,2,3†], Chulho Jung[1,3†], Junha Hwang[1,2,3], Jaeyong Shin[1,2,3], Sung Yun Lee[1,2,3], Heemin Lee[1,2,3], Seung Phil Heo[1,2,3], Daewoong Nam[3,4], Sangsoo Kim[4], Min Seok Kim[4], Kyung Sook Kim[3,4], In Tae Eom[3,4], Do Young Noh[5,6], and Changyong Song[1,2,3]*

[1]*Department of Physics, POSTECH, Pohang 37673, Korea,* [2]*Center for Ultrafast Science on Quantum Matter, Max Planck POSTECH Korea Research Initiative, Pohang 37673, Korea,* [3]*Photon Science Center, POSTECH, Pohang 37673, Korea,* [4]*Pohang Accelerator Laboratory, POSTECH, Pohang 37673, Korea,* [5]*Department of Physics and Photon Science, Gwangju Institute of Science and Technology, Gwangju 61005, Korea,* [6]*Institute for Basic Science (IBS), Daejeon 34126, Korea*

[†]*These authors contributed equally to this work.*

*Corresponding author: C.S. (cysong@postech.ac.kr)*

**Photoinduced ultrafast phenomena in materials exhibiting nonequilibrium behavior can lead to the emergence of exotic phases beyond the limits of thermodynamics, presenting opportunities for femtosecond photoexcitation. Despite extensive research, the ability to actively control quantum materials remains elusive owing to the lack of clear evidence demonstrating the explicit control of phase-changing kinetics through light–matter interactions. To address this drawback, we leveraged single-pulse time-resolved X-ray imaging of Au nanorods undergoing photoinduced melting to showcase control over the solid-to-liquid transition process through the use of localized surface plasmons. Our study uncovers transverse or longitudinal melting processes accompanied by characteristic oscillatory distortions at different laser intensities. Numerical simulations confirm that**

the localized surface plasmons, excited by polarized laser fields, dictate the melting modes through anharmonic lattice deformations. These results provide direct evidence of photoinduced surface plasmon-mediated ultrafast control of matter, establishing a foundation for the customization of material kinetics using femtosecond laser fields.

The exploration of nonequilibrium states of matter through the use of ultra-intense and ultrafast light has unveiled previously hidden phases[1-5]. By employing femtosecond (fs) photoexcitation of electrons in their ground states, highly excited states of potential energy surfaces have been effectively investigated, and the results have revealed previously unknown intermediate phases during relaxation to the ground state or metastable configurations. This approach has promoted the utilization of ultrafast photoinduced dynamics to control material properties through light–matter interactions[6-8].

A localized surface plasmon (LSP) is excited when light interacts with a metallic nanoparticle smaller than its wavelength, and it facilitates precise material control. LSPs, with their collective electron oscillations, confine light on the nanoscale, enabling active applications in sensing, imaging, energy conversion, and optoelectronics[3,9-12]. Coupling LSPs with ultrafast light may control nonequilibrium phase transition kinetics and advance quantum control of materials[5,13-15]. However, research in this field has been limited by the unexamined interactions between LSPs and changing material phases.

This study was focused on the evaluation of the role of LSPs in ultrafast photoinduced melting kinetics by imaging Au nanorods (NRs) after fs photoexcitation. The Au NRs exhibited significant near-field electromagnetic field enhancement, which was sensitive to particle orientation relative to laser polarization, providing additional control over light–matter interactions. Time-resolved X-ray free-electron laser (XFEL) experiments linked LSP modes to ultrafast photoinduced melting of single Au NRs. Numerical simulations of LSP and acoustic shape distortion modes supported the experimental findings, demonstrating LSP-controlled melting kinetics.

**Results**

We investigated the influence of LSPs on the melting transition with anharmonic lattice deformation by directly observing the ultrafast melting of Au NRs excited by a fs infrared (IR) laser pulse (800 nm wavelength, 100 fs pulse width) using time-resolved single-particle X-ray imaging. Utilizing the Pohang Accelerator Laboratory X-ray Free-Electron Laser (PAL-XFEL), we collected single-pulse single-particle data for Au NRs mounted on $Si_3N_4$ membranes by raster scanning the membrane relative to the fs-IR laser and X-ray pulse (**Fig. 1a** and **Methods**) [16-18]. The random orientation of Au NRs relative to the fs-IR laser polarization enabled the acquisition of single-particle diffraction patterns with laser polarization dependence. Different nanoparticle orientations resulted in varying near-field enhancements, leading to different effective laser fluences influenced by the enhancement factor.

Single-pulse XFEL diffraction patterns of single Au NRs photoexcited by a single fs-IR laser pulse (incident fluence of 17 $\mu$J cm$^{-2}$) showed the expansion of NRs upon melting, resulting in a shape deformation from rod to oval (Fig. 1b) [5,19,20]. The lateral size expansion exceeded the longitudinal size expansion. Numerical phase retrieval of all single-pulse diffraction patterns was performed to obtain plane-projected electron density images of the Au NRs (see Methods). Examination of these nanoscale images along three major orientations revealed significant variation in melting speed based on the orientations relative to the fs-IR laser polarization (Fig. 1c). The fastest and slowest melting occurred for NRs aligned parallel and perpendicular to the laser field, respectively, while maintaining the rod shape until 200 ps. This trend corresponds to the near-field enhancement factor of LSPs (left panel in Fig. 1c), obtained from finite-difference time-domain calculations (Supplementary Information).

For all three major orientations, melting began with void formation (arrow) near the highest electric field spots of the near-field enhancement. This void formation is due to focused plasmonic heating parallel to the laser polarization direction, causing localized ionic pressure accumulation and subsequent void creation while relieving local pressure[5]. Similar void generation was observed in single-shot imaging experiments on Au spheres and in numerical simulations[5,6,14,21]. The void then moved to the center along the longitudinal direction, eventually splitting the rod along the long axis. This expansion also caused the rod to widen along the short axis, making it oval. Apart from the reaction speed, the melting process was consistent across all three orientations, with void formation at both ends of the rod, leading to further lateral expansion and an oval shape.

Upon shape deformation to an oval upon melting, we quantified the melting rate by tracking changes in the aspect ratio (Fig. 2a and Supplementary Fig. S1). This rate was determined by fitting the temporal evolution of the aspect ratio to an exponential decay function (see Supplementary Information)[22]. The melting speed for the parallel orientation was nearly double that of the diagonal orientation and 15 times higher than the perpendicular orientation. The melting rate is proportional to the near-field enhancement factor, indicating that LSPs, excited by the enhanced near-field from fs-IR laser scattering by rod-shaped specimens smaller than the laser wavelength, influenced the melting. The maximum enhancement factor for the parallel orientation was 6.3, followed by the diagonal at 3.5, and the perpendicular at 1.2. The temporal evolution of the aspect ratio of the Au NRs in all three orientations, normalized by the near-field enhancement factor, followed the same fit curve, confirming that LSP is the primary energy source driving this ultrafast photoinduced melting (Fig. 2b).

To confirm that the melting rate was influenced by the near-field enhancement effect via LSP excitations, we repeated the experiment with increased laser fluences (23 μJ cm$^{-2}$). The melting rate, derived from the aspect ratio, maintained a linear relationship with the effective laser field (incident laser field multiplied by the near-field enhancement factor), except for the parallel orientation with a high effective field of 145 µJ cm$^{-2}$, which exhibited a different melting mode (Fig. 2b). Further experiments with an incident laser fluence of 35 µJ cm$^{-2}$ showed similar melting modes and rate slopes in parallel orientation, aligning with the results obtained at 23 µJ cm$^{-2}$.

Importantly, the temporal evolution of the long and short axis displayed an oscillatory behavior in addition to the linear expansion, as a feature distinguished from nanospheres (Fig. 2c and Supplementary Information). This oscillation was characterized with a single frequency of 41.7 GHz. To understand the underlying mechanism, numerical simulations were conducted on LSP-mode excitations, acoustic eigenmodes, and the induced stress field of the NRs (see Methods). These simulations replicated different LSP-mode excitations and resulting acoustic deformations. It was found that the oscillatory behavior aligned with the acoustic transverse shape deformation induced by the transverse plasmon excitation mode (Fig. 2d). Boundary element method (BEM) calculations revealed an Au NR surface plasmon eigenmode (left panel), with blue indicating negative and red indicating positive surface dipole strengths[23]. Finite-element method calculations demonstrated the acoustic distortion and stress field using the eigenmode model with a fixed boundary condition. The white dotted lines indicate the original shapes of the Au NRs, while the colors represent the normalized displacement and stress fields. The simulations indicate that the transverse deformation mode, driven by the fs-IR laser-excited transverse LSP mode, couples directly to induce the transverse acoustic deformation of the rod. This acoustic deformation generated a corresponding stress field distribution along the longitudinal direction, resulting in splitting along the same direction (Fig. 2d).

In contrast to the melting process observed in the low-fluence group, which involved longitudinal splitting of the rod and oval shape deformation, a distinct melting mode was identified in the high-fluence group at 23 µJ cm$^{-2}$. This process occurred in a parallel orientation, resulting in the formation of two high-density longitudinal loci resembling a dumbbell with one side larger (5 and 8 ps). Initially, the rod elongated as the two loci, marked by arrows, separated further. Over time, the distance between them reduced (10 ps), resulting in the formation of an elongated oval that approximated the original length but swelled along the short axis (20 ps). Eventually, the shape became rounder with merged high-density spots of expanded size (40 ps). A longitudinal line plot clearly illustrates this process (Fig. 3b and Supplementary Fig. S4). As melting progressed, two high-density regions emerged from 3 ps, diverging from the uniform density of the intact specimen. The distance between these regions increased until ~8 ps, then decreased, leading to the merging of the two high-density regions at ~20 ps. These separation and merging patterns were consistently observed for all the Au NRs at higher effective laser fluences (Supplementary Information).

Rod expansions along both the long and short axes were derived from all single-particle data, including those of the lower-fluence group (Fig. 3c). Compared with the long axis (expansion of ~140%), the short axis demonstrated a significantly larger expansion (~350%). As previously mentioned, melting in the high-fluence group showed an oscillatory pattern alongside linear expansion. This high-fluence melting mode with longitudinal density separation can be attributed to the photoexcitation of the longitudinal LSP mode. Numerical simulations confirmed that the oscillatory lattice expansion, with a frequency of 51.5 GHz, corresponded to the longitudinal LSP-induced longitudinal acoustic deformation (Fig. 3d and Supplementary Information). The fs-IR laser field excited the longitudinal surface plasmon mode, causing rod shape distortion with predominant longitudinal deformation. Stress-map estimation indicated local stress confinement along the longitudinal direction (Fig. 3d), which led to the splitting of two high-density loci longitudinally, consistent with experimental observations.

**Discussion**

These findings reveal that the melting process can be controlled through the excitation of LSP-coupled anharmonic acoustic deformation (Supplementary Information). The low-absorption melting process is characterized by the plasma oscillation along the short axis matching the acoustic deformation eigenfrequency (41.7 GHz), indicating the dominance of the short axis in shape distortion and melting (Fig. 2d). In contrast, high-fluence scenarios result in melting through longitudinal shape deformation (Fig. 3), with the oscillation frequency corresponding to the acoustic deformation eigenfrequency (51.5 GHz) induced by the longitudinally vibrating LSP mode. The numerical simulations conducted as part of the study showed that these two distinct melting modes are directly caused by anharmonic acoustic deformation mediated by fs-IR-photoinduced LSP modes.

The coupling between LSPs and phonons arises from ultrafast photoexcitation of electrons, resulting in energy transfer to the crystal lattice[5,21,24-27]. Initially, fs optical excitation of the LSP leads to electron-electron scattering and radiation damping, resulting in an inhomogeneous distribution of hot electrons with a higher average temperature than ions[25,28-31]. Equilibration between charge carriers and ions occurs through electron-ion scattering within a few picoseconds post-photoexcitation. The strongest coupling between LSP and phonon modes occurs when regions of the largest electric field (plasmon mode) and largest displacement (phonon mode) overlap[32]. In lower-fluence melting, the LSP mode in the transverse direction is primarily excited, driving melting through strong transverse oscillations. In contrast, higher-fluence melting predominantly excites LSP modes along the longitudinal direction, facilitating melting through strong longitudinal oscillations. This excitation induces a higher acoustic deformation mode, generating a stress field that deforms the Au NR in the longitudinal direction.

This study highlighted the impact of LSP dynamics on the ultrafast melting of Au NRs under femtosecond infrared (fs-IR) laser excitation. The Au NRs underwent distinct shape deformations influenced by the effective laser field and near-field enhancement factor. The fs-IR-laser-induced LSP directly controlled the melting modes through characteristic acoustic lattice deformations. Single-pulse time-resolved imaging at sub-10-nm and picosecond resolutions facilitated direct visualization of the LSP-controlled photoinduced melting modes, providing evidence of plasmon–ion coupling in light–matter interactions controlling solid-to-liquid phase transition kinetics. This study advances the control of material kinetics using fs laser pulses and enables quantum control of material properties through ultrafast light–matter interactions.

**Methods**

**Single-pulse single-particle time-resolved X-ray diffraction imaging**

Single-pulse single-particle X-ray imaging experiments were performed with a fixed X-ray energy of 5 keV at the Nanocrystallography and Coherent Imaging end station using the PAL-XFEL[33]. A fs Ti:sapphire laser with a wavelength of 800 nm and pulse duration of 100 fs was used as the pumping source. The laser pulses were focused to a size of 200 μm in root mean square at the position of the samples. A laser linearly polarized in the longitudinal direction was employed to photo-induce the LSPs in the Au NRs. In this study, fs X-ray laser pulses were micron focused using a pair of K-B mirrors with a focal size of 5 μm × 6 μm and a focal length of 5 m. The spatial overlap of the specimens, the 200-μm focused IR laser, and micron-focused XFEL were visually confirmed using an in-line microscope. The time zero of the interaction spot, where all three components (specimens, IR laser, and XFEL) were aligned, was confirmed by monitoring the absorption profile of the thin GaN crystal mounted on the interaction spot. The temporal resolution achieved using the PAL-XFEL was better than 0.5 ps without using a timing tool.

**Fixed-target single particle preparation for XFEL single-pulse data acquisition**

Unconjugated rod-shaped Au nanoparticles with a diameter of 50 nm and length of 145 nm, exhibiting peak surface plasmon resonance in the range of 800 nm (Nanopartz[TM]), were mounted on 100 nm thin $Si_3N_4$ membranes. The membranes were custom-designed (Silson LTD) with an array of multiple windows, forming 35 × 35 arrays with 200 μm × 200 μm windows. This design facilitated single-pulse diffraction experiments using a fixed-target sample delivery scheme[34]. To achieve a high particle hit rate, Au NRs in deionized purified water were spin-coated onto $Si_3N_4$ membranes, resulting in a nominal inter-particle distance of 3–5 μm. Single-pulse imaging experiments were conducted in the absence of a pump laser by raster scanning the sample stages.

When the pump laser was turned on, the raster-scan mode was modified to accept one pulse per window, thereby ensuring that each particle was hit by the pumping laser only once. When the $Si_3N_4$ window was exposed to the IR laser pulse, it broke completely. Laser exposure of the nanoparticles near windows was prevented by separating the window-to-window distance far more than the laser footprint. The footprint of the optical pumping laser was <200 μm, and the window-to-window distance was 380 μm. This specialized protocol ensured that the nanoparticles were exposed to the IR laser and XFEL pulses only once. Each single pulse diffraction pattern was recorded on the multiport charge-coupled device detector with a pixel size of 50 μm by 50 μm[35]. The total detector area with 1024 by 1024 was used.

**Coherent diffraction patterns, phase retrieval, and image reconstruction**

Coherent X-ray diffraction patterns from single Au NRs were obtained by exposing fresh nanoparticles to single pulses of a fs XFEL. For data analysis, we discarded the diffraction patterns of multiple-particle hits displaying interference fringes or speckles with much finer oscillation periods than expected from a single particle, and only used the data from single-particle hits. Each single-pulse diffraction pattern was phase retrieved to acquire an image of each independent nanoparticle[36-38]. This phase retrieval for acquiring a real image is not a fit process but an optimization approach to the inverse problem through numerical iterations following a phasing algorithm[16,39]. Coherent diffraction patterns were measured with a sampling interval finer than the Nyquist frequency, which was determined by the inverse sample size used for the amplitude modulus of the Fourier transformation of the specimen.

Phase retrieval was performed, starting with a randomly estimated phase for the first step. The amplitude modulus measured with this random phase was inverse Fourier transformed to render an image in real space. This image, which is generally far from an actual image of the specimen at this initial stage, was further inspected to comply with rationally imposed constraints, such as the extent of the image and the positive value in the image density. This constrained image was numerically Fourier transformed to obtain a diffraction pattern with amplitude modulus and phase. At this stage, the amplitude was replaced with the experimentally measured amplitude while maintaining the phase and inverse Fourier transform to obtain an image again. The obtained image was constrained again and Fourier transformed back into the diffraction pattern. This process was repeated back and forth and finally converged to render an image of the real specimen. As noted in this phase retrieval process, all measured data were used directly and unbiasedly for imaging, which is clearly distinguished from the data fitting process used elsewhere. Data collection was repeated for each delay time, and multiple images were obtained for the same delay time (Supplementary Video 1-8). The image that rendered the average behavior of images with the same delay time was chosen to represent the image of the given delay time.

**LSP mode calculations**

Numerical simulations of the surface plasmon modes for the Au NR were performed using the BEM, as implemented in the MNPBEM-MATLAB toolbox developed by Hohenester and Trügler[23]. The BEM simulation was initialized in quasistatic mode. The Au NR was modeled with a diameter of 50 nm and a total height of 145 nm, using 7,378 boundary elements. The complex refractive index of gold from Johnson and Christy was employed, with the surrounding environment set to vacuum (refractive index of 1)[40]. The first 20 LSP eigenmodes were computed for the Au NRs and then excited with the incident light polarized along the nanorod's long axis (Supplementary Fig. S3). Considering the amplitudes of each eigenmode, the LSP transverse modes were summed to construct the low absorption melting mode, while the LSP longitudinal modes were summed to construct the high absorption melting mode (Supplementary Information)[41].

## Displacement and stress field from acoustic distortion calculations

This simulation was performed using finite-element continuum modeling for the structural mechanics module/eigenfrequency solver in the three-dimensional finite-element software (COMSOL Multiphysics). The geometry of the Au NR was modeled as a hemispherical-capped cylinder with a diameter of 50 nm and length of ~90 nm to match the average dimensions of the Au NRs in our sample. The elastic moduli and Poisson's ratios used for the Au NR were 70 GPa and 0.44, respectively.

## Acknowledgments

This work was supported by the National Research Foundation (NRF) of Korea (grant numbers 2022M3H4A1A04074153 and RS-2024-00346711).

## Author contributions

C.S. conceived the study. E.P. and C.J. analyzed the experimental data. E.P. performed the numerical simulations and prepared single-particle specimens on fixed targets. M.K. and I.E. handled the fs-IR lasers. K.K. operated the detector. All authors contributed to the experiments. C.S. and E.P. wrote the manuscript with inputs from all authors.

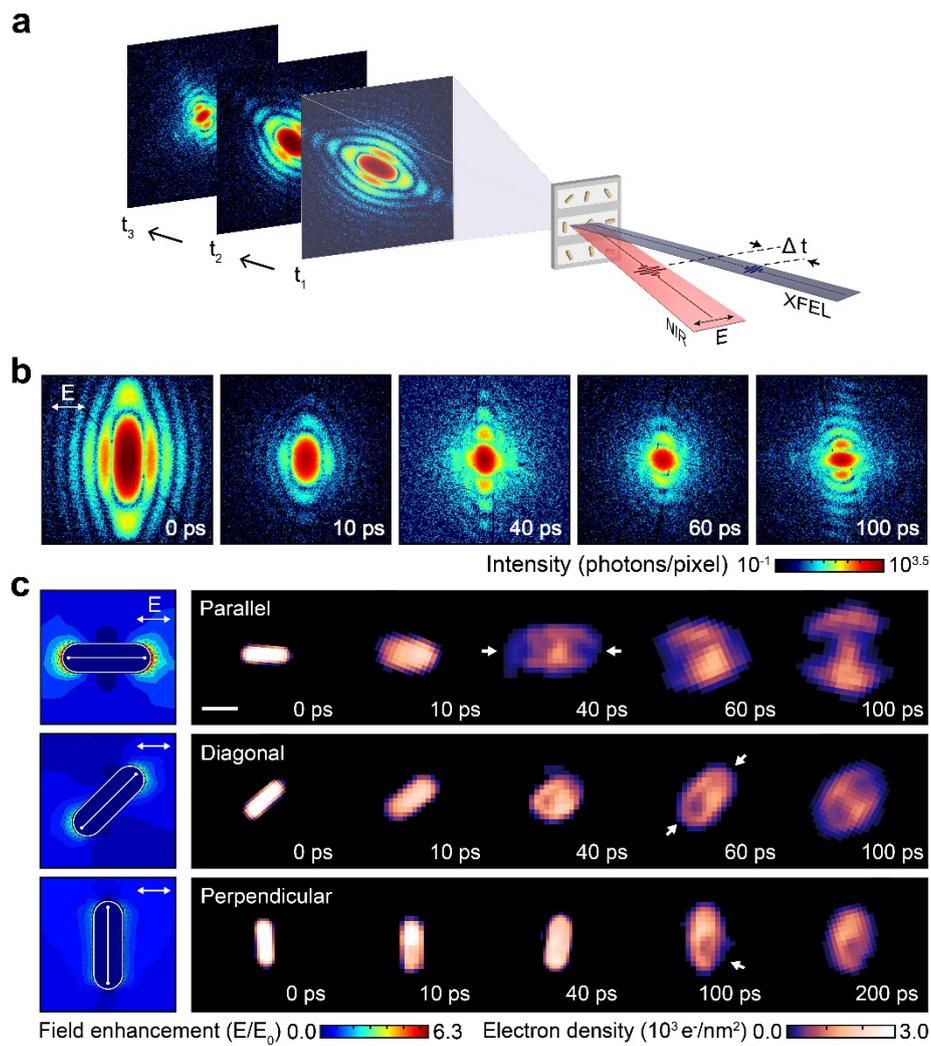

**Fig. 1: Photoinduced melting of single Au NRs with a polarization-dependent reaction speed. a**, Time-resolved single-particle X-ray imaging experiments with single XFEL pulses. Au NRs are distributed on SiN$_x$ membrane with random orientations to the incident fs-IR laser polarization. **b**, XFEL single-pulse diffraction patterns display particle melting with morphological changes from the initial rod to a deformed oval shape. **c**, Au NRs on photoinduced melting (incident laser fluence at 17 μJ cm$^{-2}$). The scale bar is 100 nm. The near-field enhancement factor is calculated (left). The melting speed depends strongly on the laser polarization direction, and the fastest melting is noted for NRs oriented parallel to the polarization with the highest near-field enhancement. Au NR melting proceeds with voids formed from both ends (arrows) of the long axis to proceed into the center to split the rod along the short axis.

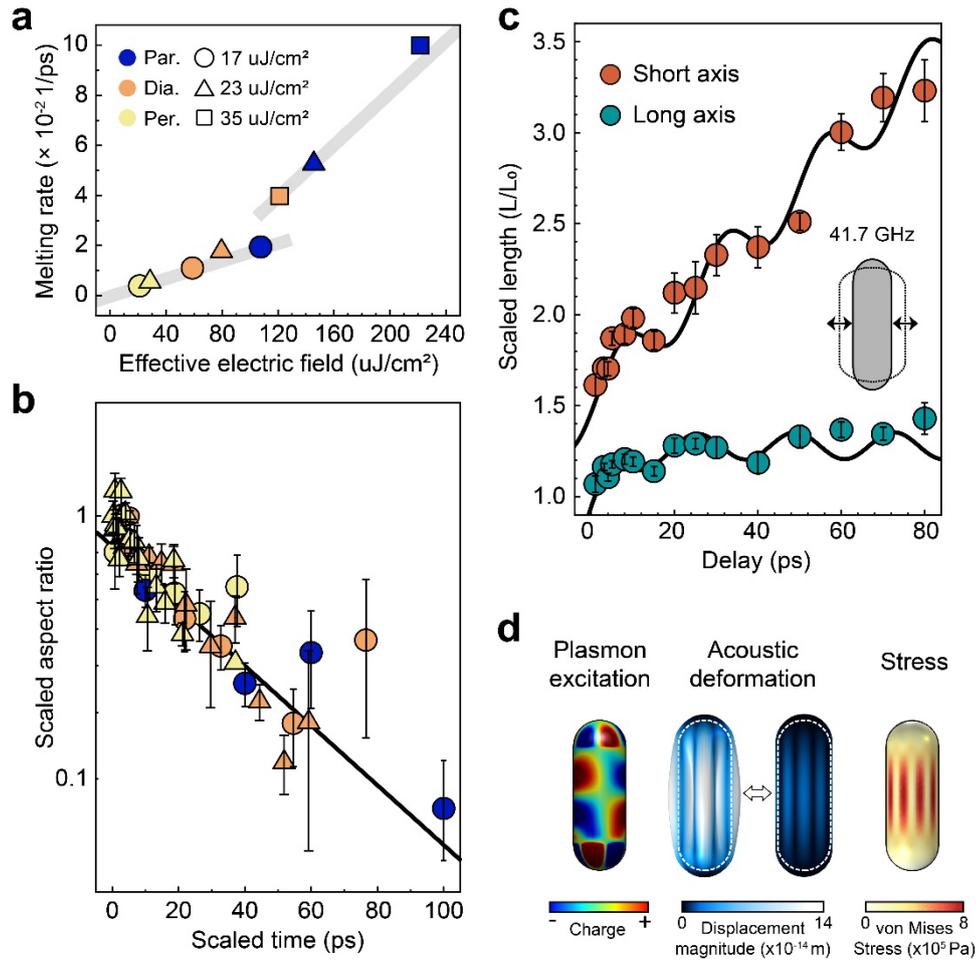

**Fig. 2: LSP-induced acoustic deformation of the NR on melting. a**, Melting rates, obtained from the aspect ratio of the rod, were extracted showing linear dependence on the effective field, multiplied by the near-field enhancement factor. Higher-fluence group showed a steeper slope. **b**, Melting reaction curves (of the lower-fluence group) merge into one exponential decay line with the reaction time linearly scaled for the effective field. **c**, Long and short axis of the rods display oscillatory expansion. The solid lines are fit with the frequency of 41.7GHz with the transverse mode (inset). **d**, Numerical simulations of the LSP mode, acoustic distortion modes, and volumetric stress fields caused by deformation of Au NRs in low-absorption mode.

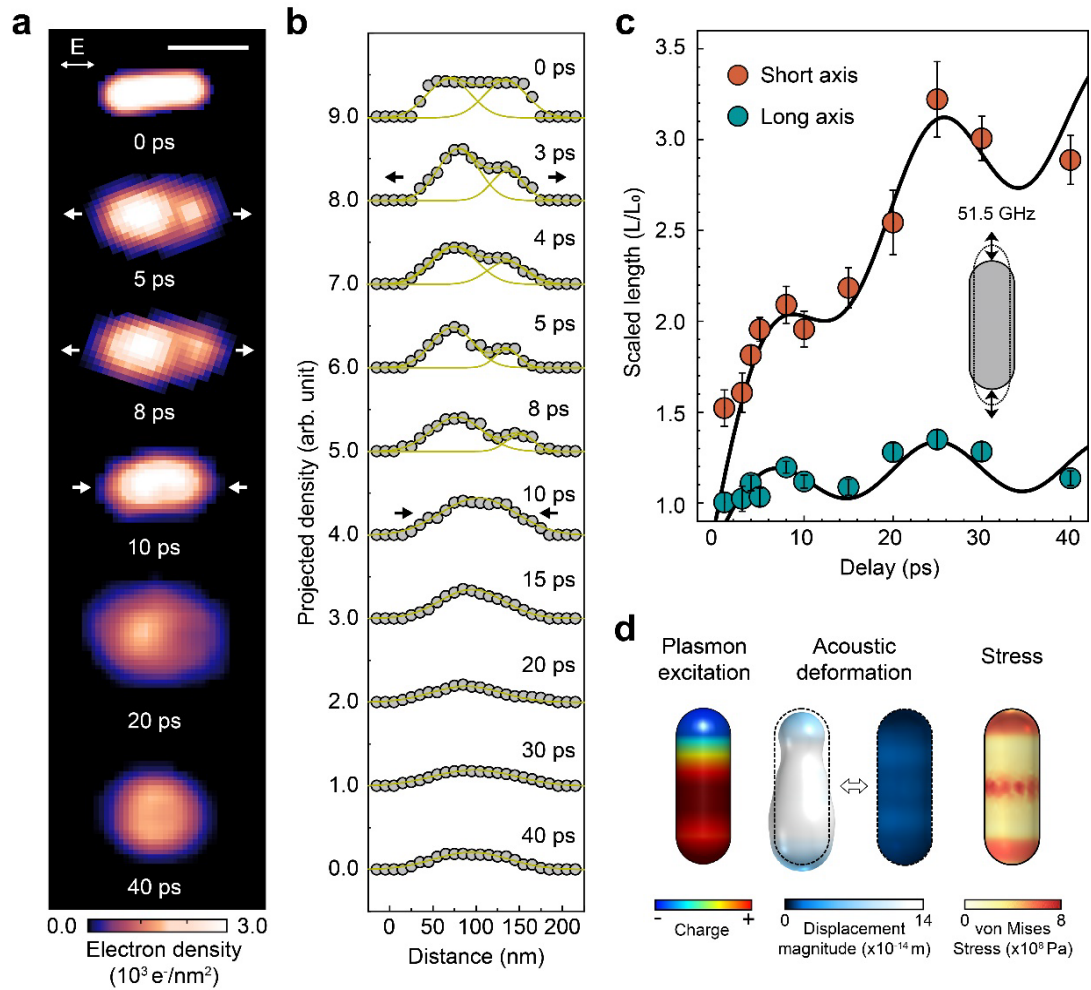

**Fig. 3: Melting in a high effective fluence with longitudinal distortion. a,** Reconstructed images display the photoinduced melting of single Au NRs for the parallel orientation in laser fluence of 23 µJ cm⁻² with the effective fluence of 145 µJ cm⁻². The melting proceeds by forming two localized high-density regions along the long axis of the NR, which eventually merge to become a circular shape. The scale bar is 100 nm. **b,** Oscillating variation of the distance between two high-density loci. Line plot along the long axis displays the process with density change. **c,** Oscillatory expansion along the long and short axis with the frequency of 51.5 GHz matching with a longitudinal vibration mode. **d,** Longitudinal LSP mode and the acoustic deformation matching the oscillation frequency of the NRs are obtained from the calculation. The stress field accumulated along the longitudinal axis leads to shape deformation of the NR upon melting.